\documentstyle[12pt,epsf]{article}

\newlength{\dinwidth}
\newlength{\dinmargin}
\begin{document}

\def\thefootnote{\fnsymbol{footnote}}

\baselineskip18pt

\thispagestyle{empty}

\begin{flushright}
\begin{tabular}{l}
  FTUAM-93/13\\\vspace*{24pt} May, 1993
\end{tabular}
\end{flushright}

\vspace*{1.5cm}

{\vbox{\centerline{{\Large{\bf VARIATIONS ON KALUZA-KLEIN COSMOLOGY
}}}}}

\vskip72pt\centerline{M.\,A.\,R. Osorio\footnote{E-mail addresses:
    {\tt OSORIO@VM1.SDI.UAM.ES} and {\tt
      OSORIO@MADRIZ1.FT.UAM.ES}.}\footnote{Address after June 1st.
    1993, Departamento de F\'{\i}sica, Facultad de Ciencias,
    Universidad de \newline\indent\hspace{5pt} Oviedo, Avda.  Calvo
    Sotelo s/n, E-33007 Oviedo, Spain (address after October 1st
    1993).} and M.\,A.  V\'azquez-Mozo\footnote{E-mail addresses:
{\tt MAVAZ@VM1.SDI.UAM.ES} and {\tt VAZQUEZ@MADRIZ1.FT.UAM.ES}.}}

\vskip12pt
\centerline{{\it Departamento de F\'{\i}sica Te\'orica
C-XI}}\vskip2pt
\centerline{{\it Universidad Aut\'onoma de Madrid}}\vskip2pt
\centerline{{\it 28049 Madrid, Spain}}

\vskip .7in

\baselineskip24pt
\indent

We investigate the cosmological consequences of having quantum fields
living in a space with compactified dimensions. We will show that the
equation of state is not modified by topological effects and so the
dynamics of the universe remains as it is in the infinite volume
limit. On the contrary the thermal history of the universe depends on
terms that are associated with having non-trivial topology. In the
conclusions we discuss some issues about the relationship between the
$c=1$ non-critical string-inspired cosmology and the result obtained
with matter given by a hot massless field in $S^{1}\times \mbox{\bf
  R}$.

\baselineskip18pt

\setcounter{page}{0}

\newpage

\section{Introduction}

As far as these authors know, the study of the influence of topology
in relativistic cosmology as for its effects on the matter filling
has
been very limited. To our knowledge, it is in the context of
Kaluza-Klein theories \cite{ACF} where this subject has been studied
for the first time by Randjbar-Daemi, Salam and Strathdee in
\cite{R-DSS}.

The Kaluza-Klein philosophy has been assumed and, in a sense,
transformed with the advent of String Theories \cite{GSW}. The
existence of extra dimensions is now a necessary ingredient.
The main reason is that after seven years of febrile activity in
the subject of String Theories, the only phenomenologically relevant
string models \cite{AGVM} are those derived from the Heterotic String
\cite{GHMR} (or of the heterotic type) which
is formulated in ten dimensions. Furthermore, the string,
being a one-dimensional object, sees the compactified dimensions in a
richer way, because of the winding states which correspond to the
wrapping of the string around the compact directions. One can believe
in a extreme
Kaluza-Klein philosophy in which all the spatial target dimensions
are
compactified at finite temperature so, in the imaginary time
formalism, everything would happen as though no open dimension
existed \cite{BV,TV}.

A quantum string is not fully apart from Quantum Field Theory. In the
canonical quantization approach we can describe the string by its
field content, i.e., with every vibrational state of the string we
associate a quantum field (analogue model). But, at the same time,
that the string is something different from a collection of quantum
fields was already evident in the old dual models because of the
$s-t$
channel duality of the four point amplitude \cite{V} (cf. also
\cite{P}). In the string models of ``everything'' this difference can
be evidentiated if we place our string in a topologically non-trivial
space, because of the property of space-time duality \cite{S} which
has no counterpart in Field Theory. In the extreme Kaluza-Klein world
described above, this implies that, assuming that the self-dual size
is the minimum size of the Universe (of the order of the Planck
scale), only those universes bigger than the Planck scale have
physical meaning. On the other hand, recent investigations
\cite{KS,OV-M1,OV-M2} have shown that at least in the large-size
regime the string is described exclusively by its field content. If
we
put together this two facts we find that String Theory would seem to
be nothing more than Field Theory plus a physical cut-off (a minimum
accessible length at the self-dual radius). In the conclusions we
will
see that this statement does not hold completely.

The final objective of our work is to investigate how the string
scenario can modify that of Kaluza-Klein presented in \cite{R-DSS} in
the extreme case described above.  However we have realized that the
subject of quantum fields in compact spaces as source for the
gravitational field is a matter of interest on its own.  Consequently
we present here, along with a set of relevant issues about the
thermodynamics of quantum fields in compact spaces, the cosmological
implications that can be extracted from the study of a two
dimensional
massless field coupled to Brans-Dicke gravity \cite{MTW} (see also
\cite{PF}) leaving the stringy modifications to be presented
elsewhere
\cite{2P}. In the conclusions we will discuss the relationship
between
a massless field in compact space at finite temperature and the $c=1$
string model as a preparation for a numerical resolution of the $c=1$
cosmology.

\section{Two-dimensional Brans-Dicke equations}

Since we are going to work in the two-dimensional case, we cannot use
Einstein-Hilbert gravity because classically it is topological in
two-dimensions. Then we are going to consider the Brans-Dicke
extension \cite{BD} which is non-trivial in any case.  The
Brans-Dicke
equations in an arbitrary number of dimensions can be written
\cite{MTW}
\begin{eqnarray}
  R_{\mu\nu}-\frac{1}{2}g_{\mu\nu}(R-2\Lambda) &=&
  \frac{8\pi}{\Phi}T_{\mu\nu}+ \frac{\omega}{\Phi^{2}}
  \left(\nabla_{\mu}\Phi\nabla_{\nu}\Phi-\frac{1}{2}g_{\mu\nu}
  \nabla_{\sigma}\Phi\nabla^{\sigma}\Phi\right) \nonumber \\ &+&
  \frac{1}{\Phi}\left(\nabla_{\mu}\nabla_{\nu}\Phi
  -g_{\mu\nu}\Box\Phi\right) \label{DBD1}\;, \\ R-2\Lambda &=&
  \frac{\omega}{\Phi^{2}}g^{\mu\nu}\nabla_{\mu}\Phi\nabla_{\nu}\Phi-
  \frac{2\omega}{\Phi}\Box\Phi\;,
\label{2BD2}
\end{eqnarray}
where $T_{\mu\nu}$ is the stress tensor for the matter fields and
$\Lambda$ is the cosmological constant.  In two dimensions the
Einstein tensor is identically zero, so the first equation simplifies
\begin{equation}
  \frac{8\pi}{\Phi}T_{\mu\nu}+ \frac{\omega}{\Phi^{2}}
  \left(\nabla_{\mu}\Phi\nabla_{\nu}\Phi-\frac{1}{2}g_{\mu\nu}
  \nabla_{\sigma}\Phi\nabla^{\sigma}\Phi\right) +
  \frac{1}{\Phi}\left(\nabla_{\mu}\nabla_{\nu}\Phi
  -g_{\mu\nu}\Box\Phi\right)= \Lambda g_{\mu\nu}\;.
\label{2BD1}
\end{equation}
For the metric we use a Friedmann-Robertson-Walker {\it ansatz}
\begin{equation}
  ds^{2}=-dt^{2}+L^{2}(t)d\xi^{2}\;,
\end{equation}
where the scale factor $L(t)$ depends only on time. We will take the
energy-momentum tensor of the matter fields to have a perfect fluid
form
\begin{equation}
  T_{\mu\nu}=(p+\rho)u_{\mu}u_{\nu}+p g_{\mu\nu}\;.
\end{equation}
$p$ and $\rho$ are respectively the pressure and the energy density.
By assuming that the Brans-Dicke field $\Phi$ is also a function of
time alone, we can rewrite (\ref{2BD2}) and (\ref{2BD1}) in the form
\begin{eqnarray}
  2\Phi^{2}\frac{\ddot{L}}{L}-2\Lambda\Phi^{2}
  &=&-\omega\dot{\Phi}^{2}+2\omega \Phi\ddot{\Phi}+2\omega
  \Phi\dot{\Phi}\frac{\dot{L}}{L}\;, \label{1.1} \\
  \dot{\Phi}^{2}-\frac{2}{\omega}\Phi\dot{\Phi}\frac{\dot{L}}{L} &=&
  -\frac{16\pi}{\omega}\Phi\rho-\frac{2\Lambda}{\omega}\Phi^{2}\;,
\label{1.2} \\
\Phi\ddot{\Phi}+\frac{1}{2}\omega\dot{\Phi}^{2} &=& -8\pi \Phi
p+\Phi^{2}\Lambda\;. \label{1.3}
\end{eqnarray}
It can be shown using (\ref{DBD1}) that the stress tensor for the
matter fields has to be covariantly conserved, so we have the
integrability condition
\begin{equation}
  \nabla_{\mu}T^{\mu\nu}=0\;.
\end{equation}
Using our {\it ansatz} for the metric this equation can be written as
\begin{equation}
  \dot{\rho}+\frac{\dot{L}}{L}(\rho+p)=0\;,
\label{1.4}
\end{equation}
which expresses the conservation of entropy in our universe. The four
equations (\ref{1.1}), (\ref{1.2}), (\ref{1.3}) and (\ref{1.4}) are
not independent. In fact, taking the time derivative in (\ref{1.2})
and combining the result with (\ref{1.3}) we see that (\ref{1.1}) and
(\ref{1.4}) are indeed equivalent. We need one equation more in order
to determine our dynamical system. This supplementary condition is
supplied by the equation of state, which relates the pressure and the
energy density $p=p(\rho,\beta,L)$.

In order to get the equation of state, we usually begin with the
Helmholtz free energy for the system, $F(\beta,L)$. All the
thermodynamics is given by this function so as to need only to
compute
derivatives of $F(\beta,L)$. In fact, we could start from the very
beginning using the Einstein-Hilbert-Brans-Dicke action
\begin{equation}
  S=\int d^{2}x \sqrt{-g} \left[\Phi(R-2\Lambda) -\frac{\omega}{\Phi}
  \nabla_{\mu}\Phi\nabla^{\mu}\Phi\right]
\label{action}
\end{equation}
and adding the matter with an action which with our {\it ansatz} can
be written as \cite{R-DSS,TV}
\begin{equation}
  S_{M}=\int dt \sqrt{-g_{00}}F\left(\beta\sqrt{-g_{00}},L\right)\;.
\end{equation}
It is easy to see that in our case the co-movil perfect fluid form of
the stress-tensor gives the same field equations as this way of
introducing hot matter.

With the {\it ansatz} for the metric, the
Hilbert-Einstein-Brans-Dicke action is invariant under the
replacement
\cite{TV}
\begin{equation}
  L\rightarrow \frac{1}{L}\;, \hspace{1cm} \Phi \rightarrow
  L^{2}\Phi\;,
\end{equation}
which corresponds to the duality symmetry of String Theory
\cite{AO2,S} upon the identification of the Brans-Dicke (dilaton)
field $\Phi$ with the inverse of the string coupling constant
squared.
To be precise, the action (\ref{action}) is invariant modulo a total
derivative
\begin{equation}
  \Delta S= 2\int dt \frac{d}{dt}\left(\Phi \dot{L}\right)\;.
\end{equation}
As a final comment, notice that the scale factor has been taken
dimensionless. Putting a length dimension in $L$ is completely
equivalent to making the change of variables $\xi \rightarrow
\xi/\lambda$ where $\lambda$ is the unit used to measure lengths. But
of course the Einstein-Hilbert-Brans-Dicke action is invariant under
this change of coordinates. In the context of String Theory
$\sqrt{\alpha^{'}}$ will play the role of $\lambda$.

When we are working with the ordinary Einstein-Hilbert action, a
non-zero vacuum energy $\Lambda_{vac}L$ can be seen as a contribution
to the cosmological constant, since in that case the perfect-fluid
energy-momentum tensor admits the decomposition
$T_{\mu\nu}=\hat{T}_{\mu\nu}+\Lambda_{vac}g_{\mu\nu}$, where
$\hat{T}_{\mu\nu}$ is the energy-momentum tensor without the vacuum
energy. Then, as can readily seen from Einstein equations,
$\Lambda_{vac}$ can be re-absorbed in a redefinition of the
cosmological constant. In the case at hand, however, we have also the
Brans-Dicke field which multiplies the energy-momentum tensor.
Nevertheless, when this field acquire a vacuum expectation value
$\langle\Phi\rangle$ we see from eqs. (\ref{DBD1}) and (\ref{2BD2})
that we can define an effective cosmological constant $\Lambda_{eff}$
given by
\begin{equation}
  \Lambda_{eff}=\Lambda-\frac{8\pi}{\langle\Phi\rangle}\Lambda_{vac}
\end{equation}
This is what one would regard as the physical (phenomenological)
value
for the cosmological constant.

\section{Thermodynamics with compactified dimensions}

We start this section computing the Helmholtz free energy for bosonic
and fermionic fields in $S^{1}\times\mbox{\bf R}$. We are going to
see
that, because of the fact that the only momentum is discrete, the
well
defined way of getting the thermodynamic potential is by computing
directly a trace on the corresponding Fock space.  In other words, as
we will explain in the conclusions, the proper time representation of
the Helmholtz free energy \cite{R-DSS} is not a well defined quantity
in this case. After all, for the time being, this representation
looks
like an unnecessary sophistication when, in the problem at hand, all
the information we are interested in can be obtained using simpler
methods.

First, let us consider a free massless scalar field.
The partition function $Z(\beta)$ is defined by
\begin{equation}
  Z(\beta)= Tr\, e^{-\beta H}\;,
\label{partition}
\end{equation}
where $H$ is the hamiltonian of the system and $\beta$ is the inverse
temperature. To directly evaluate this quantity we start by noticing
that the Fock space of a bosonic field in $S^{1}\times\mbox{\bf R}$
is
the direct product of the Hilbert space for $n$-particles. The
one-particle excitations of the system have momenta
\begin{equation}
  k=\frac{2\pi n}{L}\;,
\label{momenta}
\end{equation}
with $L$ the length of the compactified dimension and $n \in {\bf
Z}$.
We introduce creation-annihilation operators $a_{n}$, $a^{+}_{n}$
such
that the normal-ordered hamiltonian is given by
\begin{equation}
  H= {\sum_{n\in{\bf Z}}}'\;\frac{2\pi |n|}{L} a_{n}^{+}a_{n}\;.
\label{hamiltonian}
\end{equation}
The prime in (\ref{hamiltonian}) indicates that the term in the sum
with $n=0$ is omitted. This is done because the only state with $p=0$
is the vacuum state, since we are dealing with scalar massless
particles.  Using these creation-annihilation operators, we construct
the completely symmetrized states $|\{l_{n}\}\rangle$
\begin{equation}
  |\{l_{n}\}\rangle = {\prod_{n\in {\bf Z}}} '\frac{1}{\sqrt{n!}}
  (a_{n}^{+})^{l_{n}} |0\rangle\;,
\end{equation}
which span the whole Fock space. The action of the $a_{n}$,
$a_{n}^{+}$ operators on these states can be easily determined
\begin{eqnarray}
  a_{n}^{+}|\ldots,l_{n},\ldots\rangle&=&
  \sqrt{l_{n}+1}|\ldots,l_{n}+1,\ldots\rangle\;, \nonumber \\
  a_{n}|\ldots,l_{n},\ldots\rangle &=&
  \sqrt{l_{n}}|\ldots,l_{n}-1,\ldots\rangle\;.
\end{eqnarray}

Knowing the Fock space we are prepared to compute the trace in
(\ref{partition}) to give
\begin{eqnarray}
  Tr\, e^{-\beta H} &=& \sum_{\{l_{n}\}} \langle \{l_{n}\}|
  \exp{\left( -2\pi \frac{\beta}{L} {\sum_{n\in {\bf Z}}}' |n|
    a_{n}^{+}a_{n}\right)} |\{l_{n}\}\rangle \nonumber \\ &=&
  {\prod_{n\in{\bf Z}}}' \sum_{l_{n}} e^{-2\pi \frac{\beta}{L} |n|
    l_{n}}
  =e^{-\frac{\pi\beta}{6L}}\eta^{-2}\left(i\frac{\beta}{L}\right)\;,
\end{eqnarray}
where we have used the Dedekind $\eta$-function \cite{K} $
\eta(\tau)=
e^{\frac{i\pi\tau}{12}}\prod_{n=1}^{\infty}\left(1-e^{2\pi i
  n\tau}\right)$.  The Helmholtz free energy for the bosonic field
$F_{B}(\beta,L)$ is then given by
\begin{equation}
  F_{B}(\beta,L) = \frac{\pi}{6L} + \frac{2}{\beta}
  \ln{\eta\left(i\frac{\beta}{L}\right)}\;. \label{free-1}
\end{equation}
Let us remark that computing the $\beta\rightarrow\infty$ limit we
get
\begin{equation}
  F_{B}(\beta,L)\longrightarrow \frac{\pi}{6L}-\frac{\pi}{6L}=0\;.
\end{equation}
This is a simple consequence of having considered the normal-ordered
hamiltonian (\ref{hamiltonian}), since then $H|0\rangle=0$ and no
Casimir energy is present.

In order to get the equation of state, we have to obtain both the
energy density and the pressure. The first quantity is defined from
the Helmholtz free energy so as to give
\begin{equation}
  \rho(\beta,L)= \frac{1}{L}\frac{\partial}{\partial \beta}
  \left[\beta F(\beta,L)\right]=\frac{\pi}{6 L^{2}}-\frac{\pi}{6
    L^{2}} E_{2}\left(i\frac{\beta}{L}\right)\;,
\label{density}
\end{equation}
where $E_{2}(\tau)$ is a normalized Eisenstein series \cite{K}
\begin{equation}
  E_{2}(\tau)=1+\frac{6}{\pi^{2}}\sum_{m=1}^{\infty} \sum_{n\in{\bf
      Z}} \frac{1}{(m\tau+n)^{2}}= -\frac{12i}{\pi}
  \frac{\eta^{'}(\tau)}{\eta(\tau)}\;.
\end{equation}
In the case of the pressure, we have
\begin{equation}
  p(\beta,L)=-\frac{\partial}{\partial L}F(\beta,L)=\frac{\pi}{6
    L^{2}}-\frac{\pi}{6 L^{2}} E_{2}\left(i\frac{\beta}{L}\right)\;.
\label{pressure}
\end{equation}
Then, we find that our massless scalar field has the equation of
state
\begin{equation}
  p(\beta,L)=\rho(\beta,L)\;.
\end{equation}

Of course, we can reobtain the Helmholtz free energy (\ref{free-1})
using path integrals, since the partition function for a quantum
field
in a $d$-dimensional Minkowski space-time can be represented as a
Euclidean path integral for the theory in $\mbox{\bf R}^{d-1}\times
S^{1}$ with the length of the circle fixed to $\beta$ \cite{R,BL}.
The
boundary conditions we are taking for bosonic fields are periodic
along the compactified dimension ($S^{1}$) whereas for fermionic
fields antiperiodic ones are chosen in order to recover Fermi
statistics. Computing the euclidean path integral for a free scalar
field one gets that the Helmholtz free energy is given by \cite{BL}
\begin{equation}
  F_{B}(\beta)=\frac{1}{\beta}\sum_{\bf k}\ln\left(1-e^{-\beta
    \omega_{\bf k}}\right)\;,
\label{bos-PI}
\end{equation}
where $\omega_{\bf k}=\sqrt{m^{2}+{\bf k}^{2}}$ and we sum over the
momenta ${\bf k}$. In our case (a massless scalar field in
$S^{1}\times\mbox{\bf R}$) we have momenta of the form
(\ref{momenta})
and then
\begin{equation}
  F_{B}(\beta,L)=\frac{1}{\beta}{\sum_{n\in{\bf Z}}} '
  \ln{\left(1-e^{-2\pi\frac{\beta}{L}|n|} \right)}\;.
\label{3sum}
\end{equation}
We drop again the zero momentum term in the sum because, as we said,
the vacuum is the only state with zero momentum. In fact, the
inclusion of this term in the sum would produce a logarithmic
singularity. This would look like an infinite entropy (degeneration)
for the fundamental state (zero temperature limit). Introducing it
would be crooked.  It is straightforward to get (\ref{free-1}) by
converting the sum (\ref{3sum}) into an infinite product and making
use of the definition of $\eta(\tau)$.

The result (\ref{free-1}) is independent of the particular form of
the
Helmholtz free energy.  In fact, it is easy to see that as long as
$\beta F(\beta,L)$ is a function of $\beta/L$ alone
\begin{equation}
  \beta F(\beta,L)= f\left(\frac{\beta}{L}\right)\;,
\end{equation}
we can compute formally the density and the pressure with the
following result
\begin{equation}
  p(\beta,L)=\rho(\beta,L)=
  \frac{1}{L^{2}}f'\left(\frac{\beta}{L}\right)\;.
\end{equation}
So we obtain the same equation of state, namely $p=\rho$. This
functional dependence of $\beta F(\beta,L)$ is the only one possible
as long as we have no other dimensionful parameter entering in the
theory. Had we a mass term, we could have $\beta
F(\beta,L,m)=f(\beta/L, \beta m, L m)$. As we will see later the term
$L m$ does not appear for the free theory. Later, we will make some
comments about the interacting theory.

For the fermionic field we will use the result of the Euclidean path
integral computation in $\mbox{\bf R}^{d-1}\times S^{1}$ where, as
mentioned, we have to impose antiperiodic boundary conditions over
the
fermionic field along the circle $S^{1}$ of length $\beta$. In the
case we are dealing with the Helmholtz free energy turns out to be
\begin{equation}
  F_{F}(\beta,L)=-\frac{1}{\beta}{\sum_{n\in{\bf Z}}}'
  \ln{\left(1+e^{-2\pi\frac{\beta}{L}|n|} \right)}\;,
\label{fer-PI}
\end{equation}
where the zero momentum contribution has been omitted.  Again it is
quite easy to rewrite the last expression in a more manageable form
\begin{equation}
  F_{F}(\beta,L)=-\frac{\pi}{6L}-\frac{2}{\beta}
  \ln{\frac{\eta\left(i\frac{2\beta}{L}\right)}
    {\eta\left(i\frac{\beta}{L}\right)}}\;.
\end{equation}
This expression can be checked to be correct by using the relation
between the free energy for a bosonic and a fermionic free field of
the same mass \cite{O}
\begin{equation}
  F_{F}(\beta)=F_{B}(\beta)-2F_{B}(2\beta)\;,
\end{equation}
which stems from the simple mathematical identity
$(1-x)(1+x)=(1-x^{2})$ and then no mention to the proper time
representation and to any Jacobi theta function gymnastics is needed.
Since $\beta F(\beta,L)$ depends only on $\beta/L$ the equation of
state is the same as that for the boson field
\begin{equation}
  p(\beta,L)=\rho(\beta,L)=-\frac{\pi}{6L^{2}}+
  \frac{\pi}{3L^{2}}E_{2}\left(
  i\frac{2\beta}{L}\right)-\frac{\pi}{6L^{2}}
  E_{2}\left(i\frac{\beta}{L}\right)\;.
\label{dens-fer}
\end{equation}
We see that the Helmholtz free energy goes to zero when $\beta$ goes
to infinity. This is again because the hamiltonian of the fermionic
field is normal-ordered and then annihilates the vacuum state setting
the zero-point energy to zero.

It is of some interest to check that the expressions obtained so far
for the bosonic and fermionic fields recover their correct values in
the decompactification limit $L\rightarrow \infty$. First, we note
that if we naively take this limit in the expression for the free
energy of the bosonic and fermionic fields this quantity diverges.
The
reason is that in the infinite-volume limit the quantity which has
physical sense is not the total free energy but the free energy
density. We can compute the limit of this quantity using
$\eta\left(-1/\tau\right)=\sqrt{-i\tau}\eta(\tau)$.  In the case of
the bosonic field we have
\begin{equation}
  \frac{1}{L}F_{B}(\beta,L)=\frac{\pi}{6L^{2}}+\frac{1}{\beta
    L}\ln{\frac{L}{\beta}}+\frac{2}{\beta
    L}\ln{\eta\left(i\frac{L}{\beta}\right)} \longrightarrow
  -\frac{\pi}{6\beta^{2}}\;.
\end{equation}
To compute the pressure and the energy density in this limit we use
an
analogous inversion relation for $E_{2}(\tau)$ \cite{K}: $
E_{2}\left(-1/\tau\right)=\tau^{2}E_{2}(\tau)- 6i\tau/\pi$.  Applying
this formula we have the following result
\begin{equation}
  p(\beta,L)=\rho(\beta,L) \longrightarrow \frac{\pi}{6\beta^{2}}\;.
\end{equation}
For the fermionic field the situation is quite the same. The
decompactification limit of the free energy density is
\begin{equation}
  \frac{1}{L} F_{F}(\beta,L) \longrightarrow -\frac{\pi}{12\beta^{2}}
\end{equation}
and
\begin{equation}
  p(\beta,L)=\rho(\beta,L) \longrightarrow \frac{\pi}{12\beta^{2}}\;.
\end{equation}

Let us then recapitulate the results obtained so far. We find that
the
introduction of a compactified spatial dimension in our two
dimensional space-time does not modify the equation of state. On the
contrary, the dependence of the energy density and the pressure with
the temperature changes drastically. This will be important when
studying cosmological solutions of the two dimensional Brans-Dicke
equations.

In the case in which, for example, we have a compact space with
intrinsic curvature (like in \cite{R-DSS}) it is not necessary to
know
the exact form of the free energy because like in the bosonic field
in
$S^{1}\times\mbox{\bf R}$ the functional dependence on $\beta$ and
the
volume is enough to get the equation of state. For example, in the
case of a massless bosonic field in $S^{2}\times\mbox{\bf R}$ it is
easy to see that the field can be expanded in terms of spherical
harmonics
\begin{equation}
  \phi(t,\theta,\varphi)\sim \sum_{l=1}^{\infty} \sum_{m=-l}^{l}
  a_{lm} e^{i\frac{t}{R}\sqrt{l(l+1)}} Y_{\;l}^{m}(\theta,\varphi)+
  h.c.
\end{equation}
where the normalization is chosen to get the correct commutation
relations for the creation-annihilation operators and $R$ is the
radius of the sphere $S^{2}$. We see then that the momenta of the
excited states are given by
\begin{equation}
  k=\frac{\sqrt{l(l+1)}}{R}\;.
\end{equation}
Then, applying (\ref{bos-PI}), we get the Helmholtz free energy
\begin{equation}
  F_{B}(\beta,R)=\frac{1}{\beta} \sum_{l=1}^{\infty} (2l+1)
  \ln{\left(1-e^{-\frac{\beta}{R}\sqrt{l(l+1)}}\right)}\;.
\end{equation}
Since $R=\sqrt{V/4\pi}$ we can compute the equation of state by
simply
noticing that $\beta F(\beta,V)= f(\beta/\sqrt{V})$ with the result
$\rho=2 p$ as in the $(2+1)$-dimensional uncompactified case.

Moreover, we can generalize this construction to the case in which we
have a field (bosonic or fermionic) in a $ {\bf T}^{d-1} \times
\mbox{\bf R}$ where ${\bf T}^{d-1}$ is a $(d-1)$-dimensional torus
with lengths $L_{i}$. In this case the momenta of the particles are
labeled by a set of $(d-1)$ integers $(n_{1},\ldots,n_{d-1})$ such
that the momentum in the $i$-th direction is equal to
\begin{equation}
  k_{i}=\frac{2\pi n_{i}}{L_{i}}\;.
\end{equation}
By using either (\ref{bos-PI}) or (\ref{fer-PI}) we see that
\begin{equation}
  \beta F(\beta,L_{1},\ldots,L_{d-1},m)= f\left(\frac{\beta}{L_{1}},
  \ldots,\frac{\beta}{L_{d-1}},\beta m\right)\;.
\end{equation}
If we denote the total volume by $V=L_{1}\ldots L_{d-1}$ we can write
the energy density as
\begin{equation}
  \rho=\sum_{i=1}^{d-1} \frac{1}{V
    L_{i}}\partial_{i}f\left(\frac{\beta}{L_{i}},\beta m\right)+
  \frac{m}{V} \partial_{d}f\left(\frac{\beta}{L_{i}},\beta
m\right)\;,
\end{equation}
where $\partial_{i}$ indicates the partial derivative with respect to
the $i$-th entry of $f$ and $\partial_{d}$ that with respect to the
last one. The pressure in the $i$-th direction is given by
\begin{equation}
  p_{i}= \frac{1}{V L_{i}} \partial_{i} f\left(\frac{\beta}{L_{i}},
  \beta m\right)
\end{equation}
so the equation of state is
\begin{equation}
  \rho=\sum_{i=1}^{d-1} p_{i}+ \frac{m}{V}
  \partial_{d}f\left(\frac{\beta}{L_{i}},\beta m\right)\;.
\label{se-m}
\end{equation}
For an isotropical system all the $p_{i}$'s are equal, so we have
\begin{equation}
  \rho=(d-1)p+ \frac{m}{V}
  \partial_{d}f\left(\frac{\beta}{L_{i}},\beta m\right)\;.
\label{se-m-2}
\end{equation}

We are going to show that this equation of state is the same we get
in
the infinite volume limit. To see that, let us use the proper time
expression for the Helmholtz free energy \cite{P,AO} which is well
defined
\begin{equation}
  F_{B(F)}\left(\beta,V,\frac{1}{m}\right)=
  -\frac{V}{2^{\frac{d}{2}+1}\pi^{\frac{d}{2}}}\int_{0}^{\infty}
  ds\;s^{-1-\frac{d}{2}}\left[
  \theta_{3(4)}\left(0\left|\frac{i\beta^{2}}{2\pi
    s}\right.\right)-1\right] e^{-\frac{m^{2}s}{2}}\;.
\end{equation}
It is easy to check the homogeneity properties of this function
\begin{equation}
  F_{B(F)}\left(\lambda\beta,\lambda
  V,\frac{\lambda}{m}\right)=\lambda^{1-d}
  F_{B(F)}\left(\beta,V,\frac{1}{m}\right)\;.
\end{equation}
Then, applying Euler's theorem for homogeneous functions we get
\begin{equation}
  \rho = (d-1)p +\frac{m}{V}\frac{\partial F_{B(F)}}{\partial m}\;,
\label{mnoncero}
\end{equation}
which agrees with (\ref{se-m-2}). Here we have used the fact that the
free energy is a homogeneous function of degree one with respect to
the volume and then $p=-F/V$. One can also get (\ref{se-m}) by
defining the infinite volume $V= L_{1}\ldots L_{d-1}$ where each
$L_{i}$ goes to infinity. The second term in the right-hand side of
(\ref{mnoncero}) makes the equation of state for massive bosons and
fermions different from one another. Incidentally, when we set $m=0$
and $d=2$ we recover the equation of state for the massless field in
$S^{1}\times\mbox{\bf R}$. Furthermore, this is the equation of state
for a massive field in $d$-dimensions with an arbitrary number of
them
compactified.

When we deal with an intercting theory we have to proceed with more
care. Let us consider for example an interacting massless scalar
field
with dimensionless coupling constant $\lambda$ living in ${\bf
  T}^{d-1}\times \mbox{\bf R}$. In this case, because of the
divergences appearing in the Feynman diagrams containing loops, the
renormalization procedure lead us to consider an effective, scale
dependent coupling $\lambda_{eff}(\mu)$. Since we are considering the
system at finite temperature we see that the effective coupling
constant entering in our equations has to depend on both $\beta$ and
$L_{i}$, $\lambda_{eff}(\beta,L_{i})$. From dimensional arguments, we
see that
\begin{equation}
  \beta F(\beta,L_{i},\lambda_{eff})= f\left[\frac{\beta}{L_{i}},
  \lambda_{eff}(\beta,L_{i})\right]
\end{equation}
Now we can compute both the energy density and the pressure and we
get
the following equation of state
\begin{equation}
  \rho=\sum_{i=1}^{d-1}p_{i}+\frac{1}{\beta V}\left[
  \gamma_{\beta}(\beta,L_{i})+ \sum_{i=1}^{d-1}\gamma_{L_{i}}
  (\beta,L_{i}) \right]\frac{\partial f}{\partial \lambda_{eff}}
\end{equation}
where
\begin{eqnarray}
  \gamma_{\beta}(\beta,L_{i})&=&\beta
  \frac{\partial\lambda}{\partial\beta}\nonumber\\
  \gamma_{L_{i}}(\beta,L_{i})&=&L_{i} \frac{\partial\lambda}{\partial
    L_{i}}\nonumber\\
\end{eqnarray}
are a kind of {\it finite temperature beta functions}. The case of a
massive field can be treated in a similar fashion.

\section{Cosmological solutions and the thermal history}

Having all the ingredients needed, we can fulfill our other objective
which is to determine the dynamics and the thermal history of our
toy-universe. If we substitute $p=\rho$ in equations (\ref{1.2}),
(\ref{1.3}) and (\ref{1.4}) and set $\Lambda=0$ we find the following
equations
\begin{eqnarray}
  \dot{\Phi}^{2}-\frac{2}{\omega}\Phi\dot{\Phi}\frac{\dot{L}}{L} &=&
  -\frac{16\pi}{\omega}\Phi\rho\;, \label{1p} \\
  \Phi\ddot{\Phi}+\frac{1}{2}\omega\dot{\Phi}^{2} &=& -8\pi \Phi \rho
  \;, \label{2p} \\ \dot{\rho} + 2\frac{\dot{L}}{L}\rho &=& 0 \;.
  \label{3p}
\end{eqnarray}
The equation (\ref{3p}) can be easily integrated to give
\begin{equation}
  \rho(t) = \frac{\rho_{0}L_{0}^{2}}{L^{2}(t)}
\label{rho}
\end{equation}
where the subscript zero indicates the initial values for the
variables. Subtracting (\ref{1p}) from (\ref{2p}), we get a first
integral, namely,
\begin{equation}
  \frac{d}{dt}(L(t)\dot{\Phi}(t))=0
\label{ilm}
\end{equation}
so we have
\begin{equation}
  L(t)\dot{\Phi}(t)= L_{0} \dot{\Phi}_{0}\;.
\end{equation}
Using this first integral, we finally obtain a differential equation
for the Brans-Dicke field $\Phi(t)$
\begin{equation}
  -\frac{\ddot{\Phi}}{\dot{\Phi}^{2}}=\frac{\omega}{2\Phi}+
\frac{8\pi
    \rho_{0}}{\dot{\Phi}_{0}^{2}}
\label{phi-eq}
\end{equation}
that can be solved numerically. The first part of this equation can
be
related with the first derivative of the scale factor $L(t)$ to get
\begin{equation}
  \dot{L}(t)=\frac{\omega\dot{\Phi}_{0}L_{0}}{2\Phi(t)}+ \frac{8\pi
    \rho_{0}L_{0}}{\dot{\Phi}_{0}}\;.
\label{L-eq}
\end{equation}
{}From this last equation we see that whenever the Brans-Dicke field
grows big enough, the first term in the right-hand side of
(\ref{L-eq}) can be neglected and our universe will expand linearly
with slope $8\pi \rho_{0} L_{0}/\dot{\Phi}_{0}$. The range in which
this approximation is faithful depends strongly on the initial values
for the dynamical variables $\rho_{0}$, $\Phi_{0}$, $\dot{\Phi}_{0}$
and $L_{0}$. We see that, whenever $\omega<0$ we have that $L(t)$
reaches a minimum value.

To study the thermal history of our toy-universe, we make use of
equation (\ref{rho}) and the results obtained in section 3 for the
energy density of a massless field. In the case of a massless bosonic
field, we have that, according to equation (\ref{density}), the
thermal history $\beta(t)$ is given by the solution of the following
transcendental equation
\begin{equation}
  \rho_{0}L_{0}^{2}=
  -\frac{\pi}{6}E_{2}\left(i\frac{\beta(t)}{L(t)}\right)
\label{1.}
\end{equation}
where $L(t)$ is given by the solution of (\ref{L-eq}). Here we have
introduced the Casimir energy that was substracted in section 3. In
the case of a fermionic field the equation to solve is a bit more
complicated, namely
\begin{equation}
  \rho_{0}L_{0}^{2}= \frac{\pi}{3}E_{2}\left(
  i\frac{2\beta(t)}{L(t)}\right)-\frac{\pi}{6}
  E_{2}\left(i\frac{\beta(t)}{L(t)}\right)\;.
\label{2.}
\end{equation}

{}From these equations we can extract a qualitative conclusion about
the behavior of $\beta(t)$. Since the left-hand side of equations
(\ref{1.}) and (\ref{2.}) is a constant, we must have that
\begin{equation}
  \beta(t)= C_{B(F)}\times L(t)
\end{equation}
where the value of $C_{B(F)}$ is determined from the differential
equations themselves together with the thermodynamics. In the
uncompactified limit, these constants can be easily gotten from the
results of section 3, namely,
$C_{B}=\sqrt{2}C_{F}=\sqrt{\pi/(6\rho_{0}L_{0}^{2})}$.

\begin{figure}[p]
  \epsffile{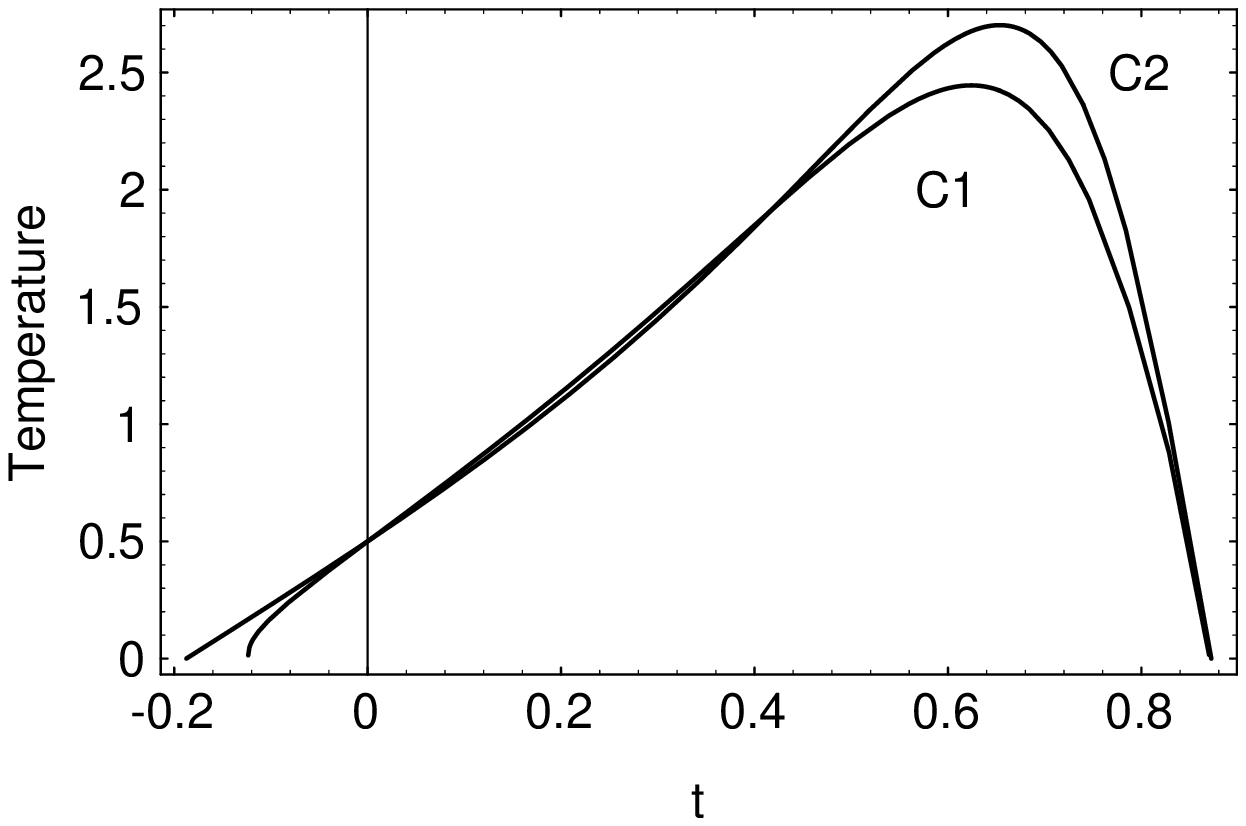}
\caption{Temperature vs. $t$ with and without a vacuum energy.}
\label{const}
\end{figure}

All this situation changes as soon as we add a vacuum energy density,
i.e., a constant to $F(\beta,L)/L$. Where this constant might come
from is an irrelevant question at this moment. Needless to say that
we
have in mind a string as a source for the matter fields in order to
justify its addition. What happens now is that the proportionality
between $\beta(t)$ and $L(t)$ is broken because of the appearance of
this constant on the right-hand side of (\ref{1.}) and (\ref{2.})
with
the corresponding signs (negative for the bosonic case and positive
in
the fermionic one). After all, the equation of state is no longer
$p=\rho$ but $p=\rho+\mbox{constant}$ so the dynamics of our universe
changes.  In fig. \ref{const} we plot the temperature vs. time for
the
situation in which the constant is absent ($C2$) and when it is
turned
on ($C1$).

\begin{figure}[p]
  \epsffile{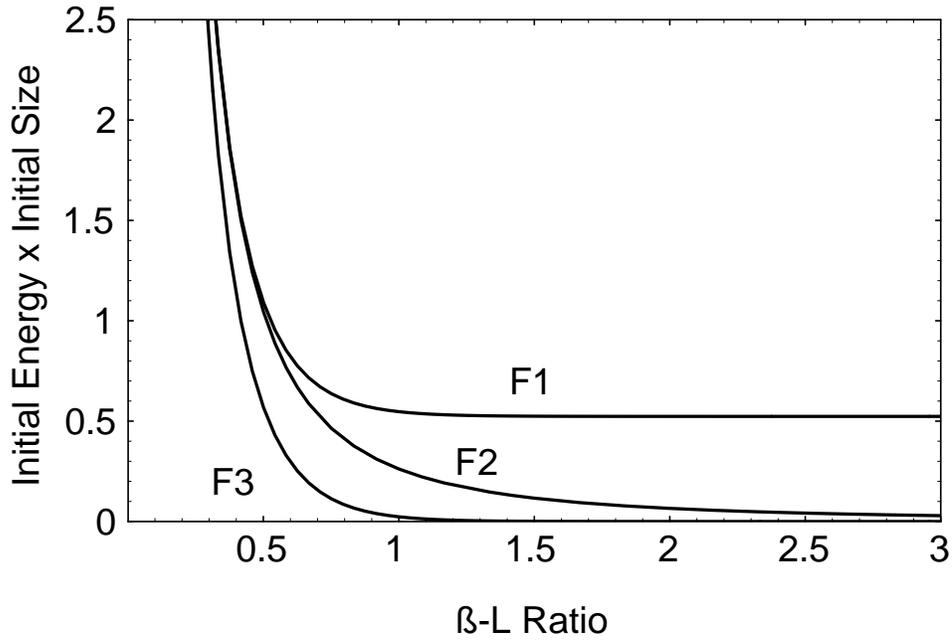}
\caption{$\rho_{0} L_{0}^2$ vs. $\beta/L$ for a massless fermionic
field.}
\label{fermi}
\end{figure}

In fig. \ref{fermi}, $\rho_{0} L_{0}^{2}$ vs. the value of the
quotient $\beta/L$ is plotted. The curve $F1$ corresponds to compact
space with a Casimir energy of $\pi/(6L^{2})$. The only effect of
this
vacuum energy is that of shifting the possible values of $\rho_{0}
L_{0}^{2}$. $F2$ is the curve for the {\it regular} $\mbox{\bf
R}^{2}$
universe at finite temperature while $F3$ is the curve for the
compactified universe after subtracting the above-mentioned Casimir
energy. Comparing $F2$ and $F3$ we notice that, since $L$ is the same
for both universes as a function of the cosmic time, the effect of
the
periodic boundary conditions in the spatial dimension is to produce,
for a given size, a universe hotter than in the open space case.

\begin{figure}[p]
  \epsffile{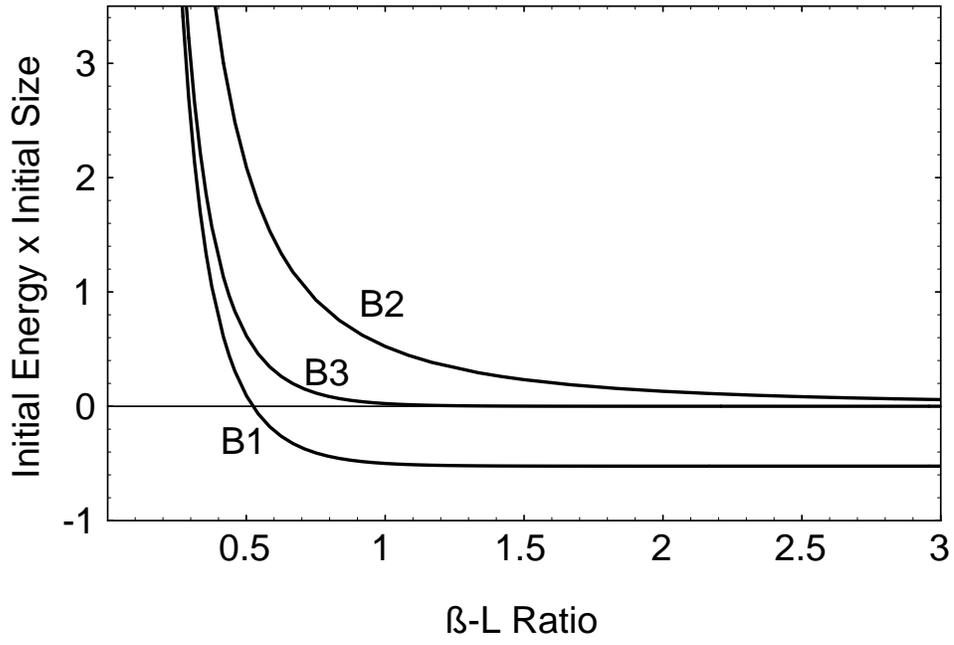}
\caption{$\rho_{0} L_{0}^{2}$ vs. $\beta/L$ for a massless bosonic
field.}
\label{bose}
\end{figure}

Essentially the same applies to the bosonic case (fig. \ref{bose}).
Now, we see that the Casimir energy is negative, so in principle we
can allow negative values for $\rho_{0}$ as it is shown in curve
$B1$.
Now we find again that the universe with the spatial dimension
compactified in a circle is hotter than its uncompactified
counterpart
($B2$) as it is evident from comparing the curve $B2$ with $B3$ in
which we have subtracted the (negative) Casimir energy.

\begin{figure}[p]
  \epsffile{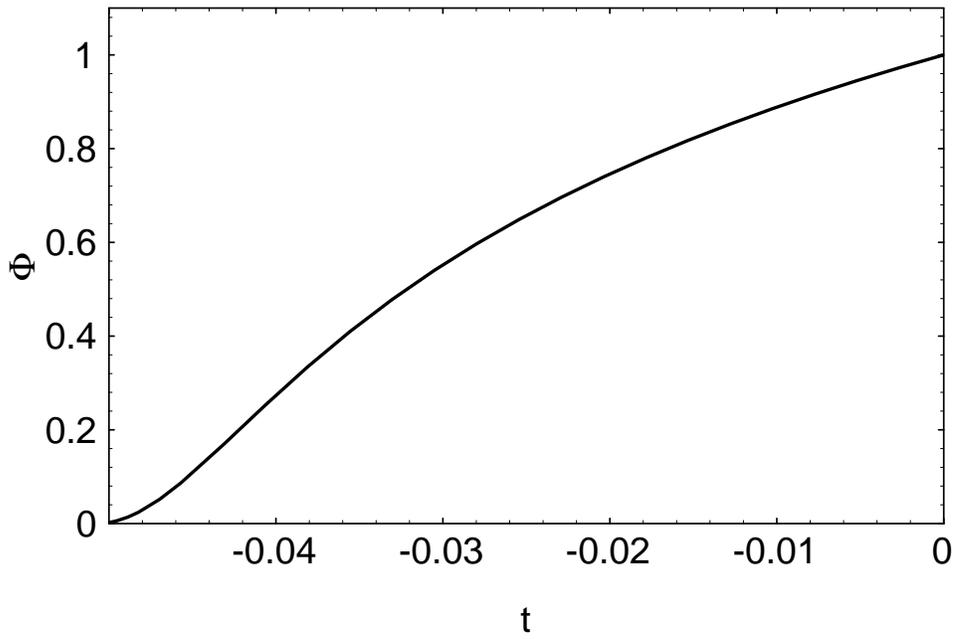}
\caption{$\Phi$ field vs. $t$ for negative time}
\label{phishort}
\end{figure}

\begin{figure}[p]
  \epsffile{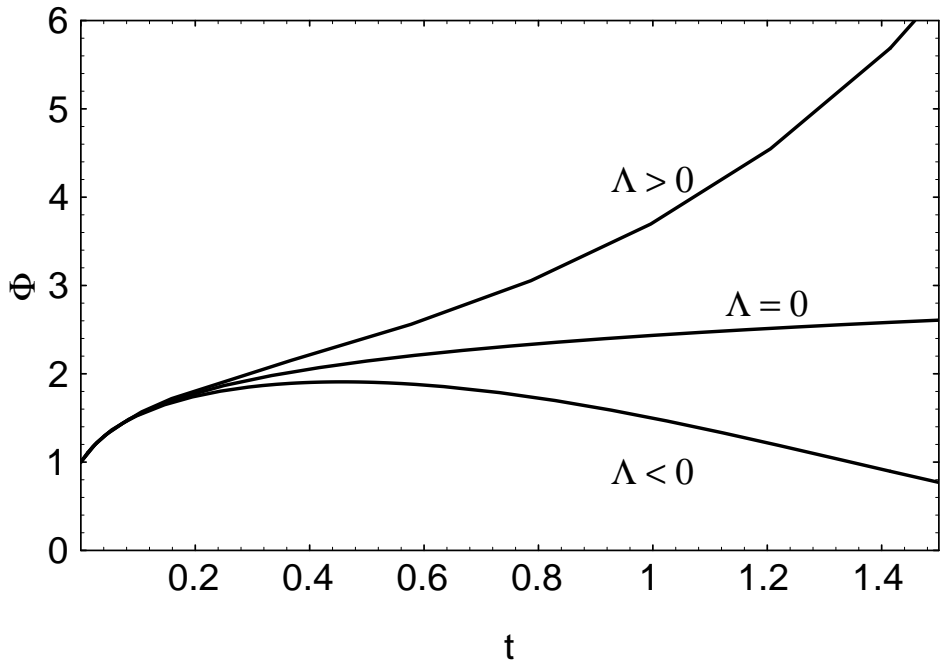}
\caption{$\Phi$ vs. $t$ for several values of the cosmological
constant}
\label{philong}
\end{figure}

When the cosmological constant $\Lambda$ is turned on, the equations
governing the dynamics are
\begin{eqnarray}
  \frac{d}{dt}\left(L \dot{\Phi}\right)&=&2 \Lambda \Phi L
\label{1.5} \\
\ddot{\Phi}+\frac{1}{2}\omega\frac{{\dot{\Phi}}^2}{\Phi} &=& -8 \pi
\rho + \Phi \Lambda\label{2.5} \\ \dot{\rho}+2\rho\frac{\dot{L}}{L}
&=&0\;. \label{3.5}
\end{eqnarray}
{}From these equations we see that, since the cosmological constant
$\Lambda$ is always multiplied by the Brans-Dicke field $\Phi(t)$,
when this is small enough the solutions are the same as those
corresponding to a vanishing cosmological constant. In the case of
equation (\ref{1.5}) we have besides the product $\Lambda\Phi L$.
This
means that, in order to get a behavior similar to the $\Lambda=0$
case
we must fulfill two conditions, namely, $\Lambda\Phi\ll 1$ and
$\Lambda\Phi L\ll 1$. This is the case for small $t$ (see fig.
\ref{phishort}) in which the curves corresponding to the $\Phi$ field
for $\Lambda>0$ and $\Lambda<0$ coincide with that of the $\Lambda=0$
case.  For the values of $t$ for which the above approximations
cannot
be made we get a splitting of the curves corresponding to the three
cases (see fig. \ref{philong}). For $\Lambda>0$ the Brans-Dicke field
grows as $\exp{(t\sqrt{2\Lambda})}$ whereas in the case of a negative
cosmological constant the $\Phi$ field reaches a maximum after which
it begins to decrease. The marginal case $\Lambda=0$ gives a
logarithmic growing for $\Phi(t)$. All the plots and the reasoning
have been made taking $\omega=-1$ having in mind the $c=1$
non-critical string (see for example \cite{GSW}).

\begin{figure}[p]
  \epsffile{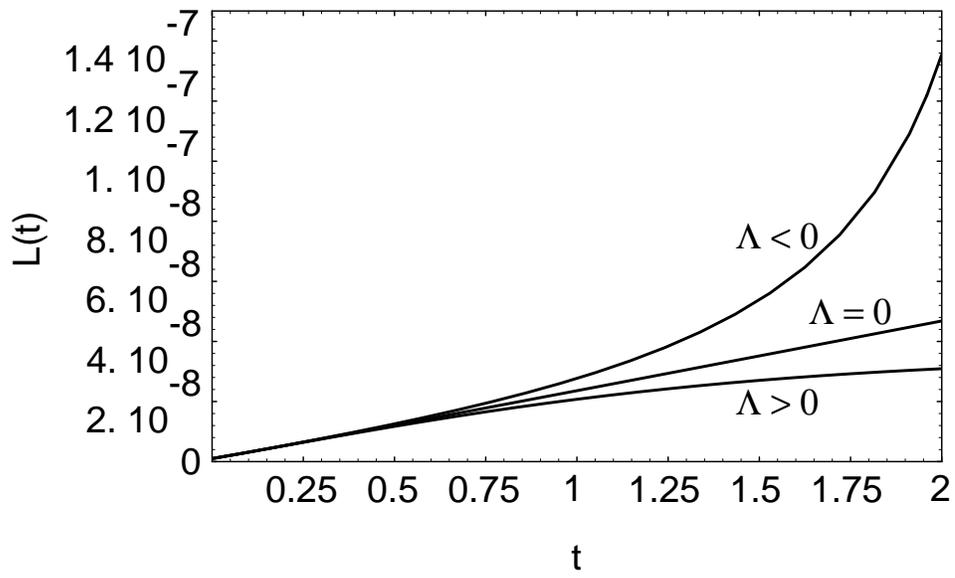} \caption{Scale factor $L(t)$ vs. $t$ for
several
    values of the cosmological constant} \label{llong} \end{figure}

The behavior of $L(t)$ can be analyzed along the lines of those for
$\Phi(t)$.  For large $t$ we get again a splitting of the curves
(fig.
\ref{llong}). The effect of a positive cosmological constant is to
produce an asymptotically static universe. On the contrary when
$\Lambda<0$ the universe inflates. For short times in which the
universe is close to its minimum length the three cases are
undistinguishable. As a final comment, it is worth noticing that the
plots have been made taking $\dot{\Phi}_{0}>0$. Owing to the fact
that
the equations are invariant under time reversal, the solutions with
$\dot{\Phi}_{0}<0$ correspond to traveling backwards in time in the
solutions plotted.

\section{Conclusions and outlook}

We have concluded that the presence of compactified dimensions has no
influence on the dynamics of the universe, since this does not modify
the equation of state of the matter. On the other hand, the
functional
relation between the pressure, the density, and the temperature
changes. As a consequence, the universe with compactified spatial
dimensions is hotter than the regular $\mbox{\bf R}^2$ universe.

A point to clarify is that of the relationship between the Helmholtz
free energy and the would-be corresponding toroidal compactification
in the massless case. In a first sight it is clear that the
relationship is broken because the would-be related compactification,
namely, the partition function for a massless particle on a torus,
should be invariant under the exchange $\beta \leftrightarrow L_{i}$.
A bridge joining both calculations would be the proper time
representation of the Helmholtz free energy. The main question is
whether that representation exists in this case and is well defined.
It is easy to apply the expression given in \cite{R-DSS} for $\beta
F(\beta)$ to the case of $S^{1}\times\mbox{\bf R}$ (this is easily
generalizable to any number of compactified dimensions) to get
\begin{eqnarray}
  \beta F(\beta)&=& L\left.\frac{d}{dt}\left(\frac{1}{2
    \Gamma(-t)}\int_{0}^{\infty} \; ds \; s^{-t-1}\, e^{-s m^2}
  \theta_{3}\left( 0\left|\frac{4\pi^2 s}{\beta^2}\right.\right)
  \theta_{3}\left(0\left| \frac{4\pi^2 s}{L^2}\right.
\right)\right)\right|_{t=0}\nonumber \\ &=&
-\frac{1}{2}\int_{0}^{\infty} \frac{ds}{s} \;\theta_{3}\left(
0\left|\frac{2\pi^2 s}{\beta^2}\right.\right)
\theta_{3}\left(0\left|\frac{2\pi^2 s}{L^2}\right.\right)e^{-s m^2/2}
\;, \label{m}
\end{eqnarray}
where the second equation follows from the first one by formally
taking the derivative. Both integrals are ultraviolet divergent. We
can fix the problem by subtracting the vacuum energy corresponding to
the limit in which $\beta\rightarrow \infty$ and $L\rightarrow
\infty$
simultaneously, i.e.,
\begin{equation}
  -\frac{L\beta}{4\pi}\int_{0}^{\infty} \frac{ds}{s^2} e^{-s m^2/2}
\label{subst}
\end{equation}
Setting $m=0$ in (\ref{m}) we add another divergence in the
$s\rightarrow \infty$ limit. Being only concerned about mathematics
one can fix again the problem by subtracting at the same time
\begin{equation}
  -\frac{1}{2}\int_{0}^{\infty} \frac{ds}{s}
\label{m=0int}
\end{equation}
Physically, this divergence is the result of including the
second-quantized version of the zero momentum state that, when
computing the trace, is suppressed because, when $m=0$, the only
state
with zero momentum is the vacuum state. If we try to keep on track of
this expression by commuting the integral with the sums in order to
extract the regulated finite result, we find that the output of this
manipulation does not depend on both $\beta$ and $L$.  The situation
changes when an ultraviolet cutoff in proper time is included in the
second part of (\ref{m}) with $m=0$ because in that case the formal
integration produces an expression which at least depends on $\beta$
and $L$. For this phenomenon to occur it is crucial the form of the
integration measure, $ds/s$, since there is no jacobian induced by a
linear change of variables and, in the absence of a cutoff, the
limits
of integration are not modified either. Using this ultraviolet lower
limit for the proper time we still have to subtract the infrared
divergence.  Instead of putting a cutoff we can also try dimensional
regularization, and at the same time subtract (\ref{subst}) from
(\ref{m}) with $m=0$ to fix the ultraviolet divergence. Doing so we
get
\begin{equation}
  \beta F(\beta,L)=\ln{(\mu L)}+
2\ln{\left[\eta\left(i\frac{L}{\beta}
  \right)\right]}
\label{regboson}
\end{equation}
where we have subtracted the term in $1/(d-1)$ together with other
constant terms and $\mu$ is the energy scale introduced when
dimensionally regularizing. This expression enjoys the invariance
under the exchange $\beta\leftrightarrow L$ (although it does not
give
the partition function of a massless boson in $S^{1}\times S^{1}$).
We
are now very close to a string variation of our massless field. If in
the right hand side of (\ref{regboson}) we make the replacement
$\beta\rightarrow (2\pi)^2/(\mu^2\beta)$ and add the resulting new
term to (\ref{regboson}) we get, upon the identification
$\sqrt{\alpha^{'}}=\mu^{-1}$, what would be the one-loop free energy
for the $c=1$ model in a $S^{1}\times \mbox{\bf R}$ target
\cite{DKL,OV,TV}. By construction this new expression in invariant
under the replacements $\beta\leftrightarrow L$ and the
transformation
$\beta \rightarrow (2\pi)^2/(\mu^2\beta)$. An inmediate consequence
is
that the equation of state is no longer $p=\rho$ \cite{2P}. After
all,
strings are not equivalent to quantum fields. It seems that they are
equivalent to fields with a kind of ultraviolet cutoff but not at the
self-dual point; what we have is a cutoff in proper time provided by
modular invariance. However, the Helmholtz free energy gotten from
the
dimensionally regularized massless boson using this procedure
presents
unphysical thermodynamical properties. For example it gives a
negative
infinite entropy in the low temperature limit. This lead us to claim
that the equivalence between the Helmholtz free energy and the
toroidal compactification is also broken for the $c=1$ non-critical
string \cite{2P}.

\section*{Acknowledgements}

We thank E. Alvarez for reading the manuscript and making useful
suggestions. Interesting suggestions were also made by J\,.L\,.F\,.
Barb\'on and A. Nieto. We survived our first contact with the {\it
  Mathematica} package thanks to the valuable help of G.
Mart\'{\i}nez-Pinedo. This work has been partially supported by CICyT
project No. AEN90-0272.  M.A.V.-M. is a Comunidad Aut\'onoma de
Madrid
Predoctoral Fellow.

\newpage

\begin{figure}[p]
  \epsffile{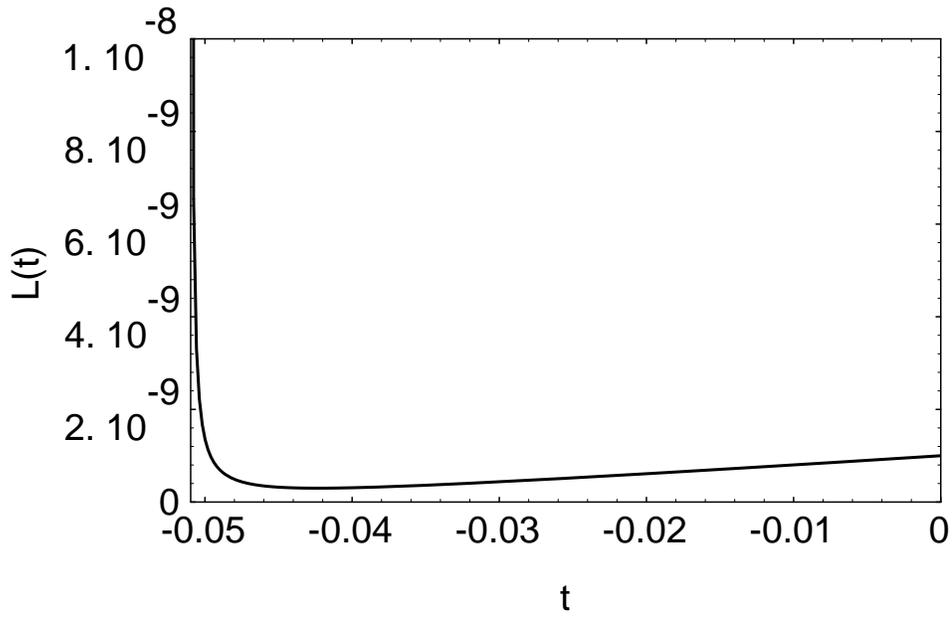}
\caption{$L(t)$ vs. $t$ for negative time}
\label{lshort}
\end{figure}

\end{document}